\documentclass[entropy,article,accept,moreauthors,pdftex]{mdpi}
 
\RequirePackage{color}
\definecolor{MyDarkGreen}{rgb}{0.02,0.60,0.06}

\firstpage{1}
\pubvolume{1}
\issuenum{1}
\articlenumber{0}
\pubyear{2021}
\copyrightyear{2021}
\externaleditor{Academic Editor: {Firstname Lastname}} 
\datereceived{}
\dateaccepted{}
\datepublished{}
\hreflink{https://doi.org/} 
 
\Title{Generalized Ising Model on a Scale-Free Network: An Interplay of  Power Laws}

\TitleCitation{Generalized Ising Model on a Scale-Free Network: An Interplay of  Power Laws}

\Author{Mariana Krasnytska $^{1,2,*}$\orcidA{}, Bertrand Berche $^{2,3}$ \orcidB{}, Yurij Holovatch $^{1,2,4}$\orcidC{} and Ralph Kenna $^{2,4}$\orcidD{}} 

\AuthorNames{Mariana Krasnytska, Bertrand Berche, Yurij Holovatch and Ralph Kenna}

\AuthorCitation{Krasnytska, M.; Berche, B.; Holovatch, {Yu.}; Kenna, R.} 

\address{%
$^{1}$ \quad Institute for Condensed Matter Physics, National Academy of Sciences of Ukraine, {UA-79011} 
 Lviv, Ukraine; hol@icmp.lviv.ua\\
$^{2}$ \quad ${\mathbb L}^4$ Collaboration \& Doctoral College for the
Statistical Physics of Complex Systems,
Leipzig-Lorraine-Lviv-Coventry 
\\
$^{3}$ \quad Laboratoire de Physique et Chimie Th\'eoriques, Universit\'e de
Lorraine, BP 70239, {CEDEX,}  54506~Vand\oe uvre-les-Nancy, France; {bertrand.berche@univ-lorraine.fr}\\ 
$^{4}$ \quad Centre for Fluid and Complex Systems, Coventry University, Coventry
CV1 5FB, UK; {r.kenna@coventry.ac.uk} }

\corres{Correspondence: kras.marjana@gmail.com}

\abstract{We consider a recently introduced generalization of the Ising model in which individual 
spin strength can vary.
The model is intended for analysis of ordering in systems comprising agents which, although matching in their binarity (i.e., maintaining the iconic Ising features of `+' or `$-$', `up' or `down', `yes' or `no'), differ in their strength. 
To investigate the interplay between variable properties of nodes and interactions between them, we study the model on a complex network where both the spin strength and degree distributions are governed by power laws. We show that in the annealed network approximation, thermodynamic functions of the model are self-averaging and we obtain an exact solution for the partition function. 
This allows us derive the leading temperature and field dependencies of thermodynamic functions, their critical behavior, and logarithmic corrections at the interface of different phases.
We find the delicate interplay of the two power laws leads to new universality classes.
}
 
\keyword{Ising model; scale-free network; self-averaging; steepest descent}

\begin{document}

\section{Introduction}\label{I}

It is almost  futile, and perhaps impossible, to  comprehensively list the advances in understanding of various phenomena in physics and beyond that were achieved due to the Ising model. 
 Excellent reviews of the one-hundred year history of the model \cite{reviews1,reviews2,reviews3,reviews4,reviews5,reviews6} are supplemented by discussions in other papers of this Special{ Issue.} 
 {This paper} has been written for the Special Issue of {\em Entropy} '{\em Ising Model: Recent 
Developments and Exotic Applications}'. 
We think it is  therefore more beneficial to  open our paper with two first-hand accounts that concern Ernst Ising, the person and the model. The first of these is of a historical nature and concerns another body of work by the 
present authors and their colleagues. 
The second, rather methodological account, will bring us closer to the subject of studies of new physics presented  in this paper.

For a quarter of a century, the {\em Ising lectures} have facilitated the emergence of different initiatives, both spontaneously and by design, that both review and advance Ising model-related research \cite{Isinglectures}.
This workshop started in Lviv (Ukraine) in 1997 with 'traditional' statistical physics  and has recently broadened its scope to encompass a more general context of complex systems. 
The lectures  became the subject of a review series \cite{Holovatch04_20_1,Holovatch04_20_2,Holovatch04_20_3,Holovatch04_20_4,Holovatch04_20_5,Holovatch04_20_6} and gradually the workshop gave rise to various research projects centered around the Ising model and its history. 
Historical documents collected to date, and displayed publicly with permission of Ernst Ising's family, include his dissertation \cite{Ising24a} and its shortened version which was published in Hamburg in 1924 \cite{Ising24b}.
They also include memoirs of  Ernst's wife, Johanna (Jane) Ising \cite{Jane}, as well as a recent publication that includes memoirs of their son Thomas \cite{Ising17}. 
It was through this collaborative atmosphere of the workshop, and in the  context of a broader ${\mathbb L}^4$ Collaboration in Statistical Physics of Complex Systems  \cite{L4}, that the problem considered below emerged. 

As mentioned, the second remark brings us closer to the scientific subject of this paper; it concerns a special feature which made the Ising model so popular for descriptions of collective behavior in multitudes of systems. 
In its original form, as presented in Ising's thesis, this feature is binarity---representation of the state of an agent as from a pair of binary oppositions.
It is to a large extent due to this feature that the model has been (and we believe will continue to be) applied in almost all fields where binarity plays a core role \cite{Ising17,Stauffer00,Holovatch17}. 
Some generalizations of the Ising model lose this feature. 
An example is the $q$-state Potts model \cite{Potts52,Wu82} which keeps the discrete symmetry of the Ising model, generalizing it from $Z_2$ to $Z_q$.
As a result, although each agent (spin) can take on only a finite number of states, the binarity is lost for any $q\neq 2$. 
Another popular generalization, the $O(m)$-symmetrical model \cite{Stanley68,Stanley71}, enables an infinite number of states  for a single agent because the symmetry is continuous at $m\neq 1$. 

Here, we address ordering phenomena in systems of agents that are not necessarily physical in nature with the special role that is played by spin models in complex networks in mind \cite{Holovatch17,Dorogovtsev08}. 
Recently, we have suggested another generalization of the Ising model that tackles such circumstances by keeping binarity of the Ising model but relaxing the condition of fixed spin length on each site \cite{Krasnytska20}. 
Within the model, the length of each spin is considered as a quenched random variable with a given distribution function and hence the observables  are calculated by the usual Gibbs averaging over the (up and down) spin configurations as well as over the random spin length distribution. 
The model is related to (but differs from) other spin models that are used to study the impact of structural disorder on collective behavior \cite{Mattis1,Mattis2,Hopfield1,Hopfield2,Hopfield3,spin_glasses1,spin_glasses2,Folk03} and it may be useful in analysis of ordering in magnetic or ferroelectric systems of particles with polydisperse elementary moments~\cite{Tadic1,Tadic2}. 
Another obvious field of applicability of this model is understanding peculiarities of ordering processes in systems containing agents that, although being of binary character (`+' or `$-$', `up' or `down', `yes' or `no'), differ in strength of expression~\cite{Galam12,Holyst17}.

\begin{figure}[H]
{\includegraphics[width=13cm]{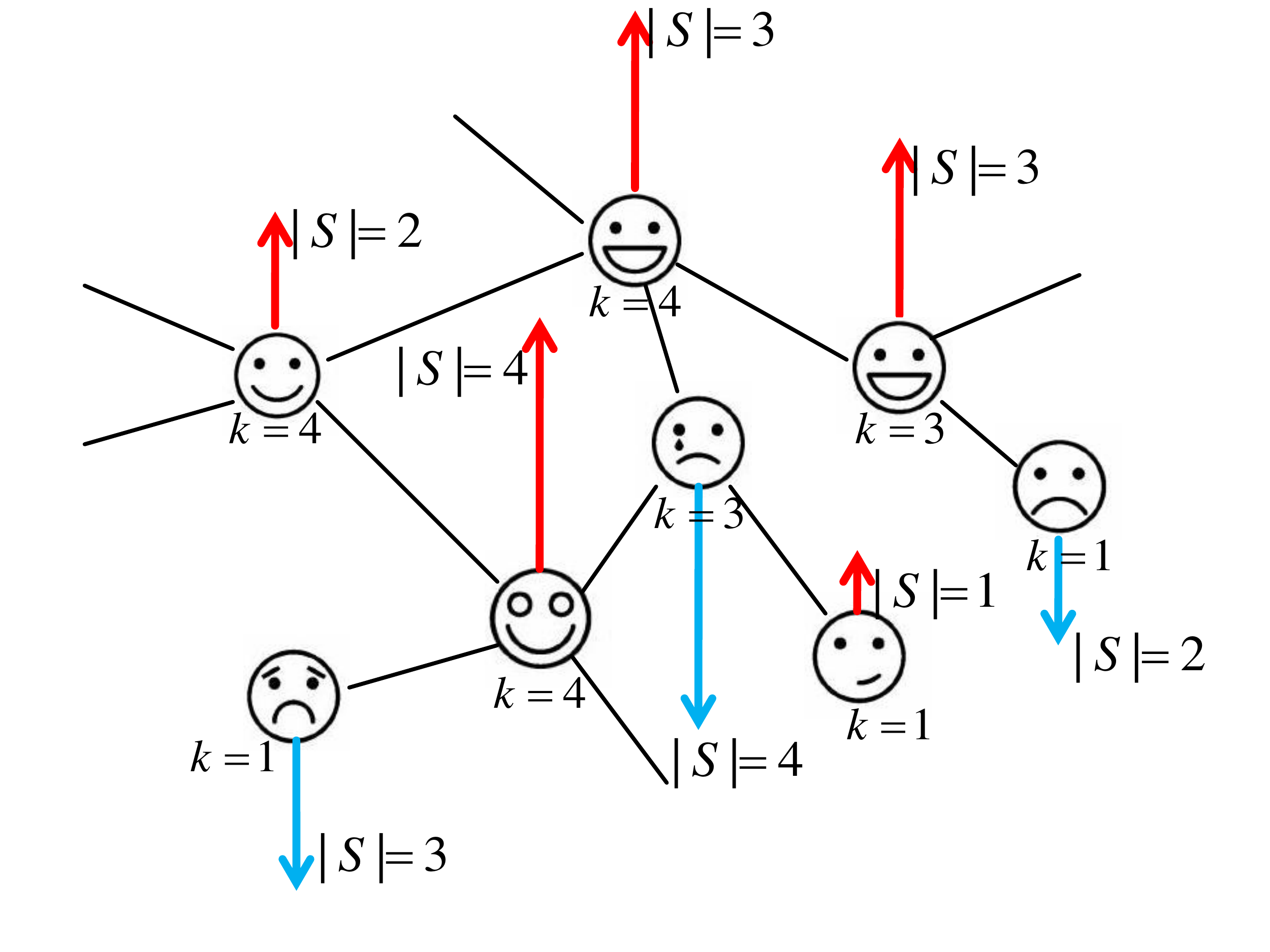}}

\caption{Ising model with varying spin length (strength) as a model for
a social phenomenon. Each individual is represented as a complex network
node of a given degree $k_i$ (i.e.,~a number of persons 
connected to it via social links) and given strength $\mathcal{S}_i$. One may 
consider spreading of positive 
(spins up)  and negative (spins down) emotions in a social network.
   \label{fig1}}
\end{figure}

An example is illustrated in Figure~\ref{fig1}. 
The structure of the  network is used to model the underlying interactions in a system of interest, be they of specific chemical, biological, social, or economic origin.
In a recent short communication \cite{Krasnytska20}, we reported on  the peculiarities of the generalized Ising model when the random spin length is governed by  a  power-law decaying distribution function. 
We obtained an exact solution for this model on complete and Erd\H{o}s-R\'eny graphs as well as commented on the phase diagram of this model on an annealed  scale-free network. 
The analytic solution for this last case has never been displayed to date and is a subject of this paper.
The  rest of the paper is organized as follows. In Section \ref{II}, we formulate
the model and demonstrate that the partition function of the model possesses
an important feature: it is self-averaging. This fact essentially facilitates calculations 
of thermodynamic functions as displayed in Section \ref{IV}. We apply the steepest 
descent method to get exact results on the thermodynamic limit. We also analyze the phase
diagram and show how an interplay between two different power laws, one governing the
network structure and another one governing spin properties, defines universal
features of critical behavior. Conclusions and outlook are given in Section \ref{V} 
and asymptotic estimates for the
integrals that enter thermodynamic functions are derived in Appendix~\ref{app1}.
\section{Model}\label{II}

Well-studied generalizations of the Ising model include the $m$-vector \cite{Stanley68,Stanley71} and the Potts \cite{Potts52,Wu82} model.
Instead of a discreet scalar variable $\sigma_i=\pm 1$, the former considers a classical vector variable $\vec{\sigma}_i$ that can point in any direction in an $m$-dimensional space.
The Potts model, on the other hand, maintains discrete variables, but relaxes the number of single-site spin states. 
Here, we consider another generalization of the Ising model. 
The new model preserves the binary character of the spin variables but allows
them to change their absolute value in a continuous and random manner \cite{Krasnytska20}. To achieve this, we endow the spins with `strength' that can vary through a random variable $\cal{S}$ with a given probability distribution function $q(\mathcal{S})$. 
Below, we consider the case where this  distribution function is characterized by a power-law decay:
\begin{equation}\label{2.1}
q(\mathcal{S})=c_{\mu}\mathcal{S}^{-\mu}, \hspace{3em} \mathcal{S}_{\rm min} \leq 
\mathcal{S} \leq \mathcal{S}_{\rm max},
\end{equation}
\textls[-15]{with the normalization constant $c_\mu$ and $\mu>2$ to ensure finiteness of the mean strength $\langle \mathcal{S} \rangle$  at $\mathcal{S}_{\rm max} \to \infty$.
As mentioned in the Introduction, the model mimics inhomogeneities in many-particle (multi-agent) systems of different natures, that may range from polydisperse magnets or ferroelectrics \cite{Mattis1,Mattis2,Hopfield1,Hopfield2,Hopfield3,spin_glasses1,spin_glasses2,Folk03,Tadic1,Tadic2} to various complex social or economical
systems \cite{Galam12,Holyst17}. 
In turn, the choice of the distribution function in the form of a power law allows both to proceed with analytic calculations as well as to gain access to various regimes of polydispersity by tuning exponent $\mu$.}

Considering the critical behavior of a spin system on a complex network, special attention has been paid to  scale-free networks, which are characterized by a power-law decay of a node degree distribution function:
\begin{equation}\label{2.2}
p(K)=c_\lambda K^{-\lambda}, \hspace{3em} K_{\rm min} \leq 
K \leq K_{\rm max},
\end{equation}
where $p(K)$ is the probability that any given node has degree (number of links) $K$, $c_\lambda$ is a normalization constant, and $\lambda>2$. 
It is well  established by now that the Ising model on a scale-free network has a
non-trivial critical behavior: depending on the value of $\lambda$, it is characterized by different critical exponents  \cite{Leone02,Goltsev03,Palchykov09}. 
For example, when $\lambda>5$,  the critical exponents coincide with the mean-field ones observed for regular lattices. 
In the region $3<\lambda<5$, the exponents become $\lambda$ dependent. When $\lambda=5$,  logarithmic corrections to scaling appear.

Below, we consider a generalized Ising model with varying spin strength on a scale-free network. 
Doing so, we analyze how an interplay of power laws (\ref{2.1}) and (\ref{2.2})---the first governing network structure and the second governing agents' strengths---impacts  critical behavior. 
To proceed, we first formulate the annealed network approximation we will be dealing with.

\subsection{Ising Model on an Annealed Network}\label{II.1}

Following Refs. \cite{Lee09,Bianconi12,Krasnytska15,Krasnytska16}, we define an annealed
network as an ensemble of networks of $N$ nodes each, with a given
degree {arrangement} $\{K\}=(K_1, K_2, ..., K_N)$, maximally random under the constraint that their degree distribution is a given one. 
The linkage between nodes is taken to fluctuate for each fixed sequence $\{K\}$. 
Therefore, in the spirit of the concept of annealed disorder \cite{Brout59}, the partition function is to be averaged with respect to these fluctuations. 
This is different from quenched disorder, when for each fixed sequence $\{K\}$ network links are fixed too and therefore the free energy is averaged. 
In this latter case, the configurational model serves as a counterpart of the annealed
network (see, e.g.,~\cite{Dorogovtsev02}).

To construct an annealed network of $N$ nodes, one assigns to each node $i$ a random variable (label) $k_i$ taken from the distribution $p(k)$ and the probability of a link between two nodes is defined as:
\begin{equation}\label{1}
p_{ij}= \frac{k_ik_j}{N\langle k \rangle}+O(1/N^2)\, ,
\end{equation}
with $\langle k \rangle=\frac{1}{N}\sum_lk_l$. 
One can show that the value of the random variable $k_i$ indicates the expected value of the node
degree: $\mathbb{E} K_i=\sum_j p_{ij}=k_i$ whereas its distribution $p(k)$ defines node degree
distribution $p(K)$.

In the presence of a homogeneous external magnetic field $H$, the Hamiltonian of the (usual) Ising model on an annealed network reads:
\begin{equation}\label{2a}
{\cal H} = - \frac{1}{2}\sum_{i\neq j} J_{ij} \sigma_i \sigma_j - H
\sum_i \sigma_i \, , \hspace{1cm} \sigma_i=\pm 1,
\end{equation}
where the second sum spans all  $N$  network nodes, the first is
over all their pairs and $J_{ij}$ is an adjacency matrix with matrix
elements equal to $J$ if nodes are connected and $0$ otherwise:
\begin{eqnarray}\label{3}
 J_{ij}=\left\{
\begin{array}{ccc}
                 J,& \hspace{0.5em} p_{ij}\, ,\\
                 0,& \hspace{0.5em} 1- p_{ij}\, .
              \end{array}
  \right.
\end{eqnarray}
For the fixed  sequence of random variables $\{k\}=(k_1,......,k_N)$, the partition
function is obtained by averaging with respect to random annealed links $\{J\}$:
\begin{equation}\label{4}
{\cal Z}_N(\{k\}) = \langle {\rm Sp}_{\sigma} e^{-\beta {\cal H}}
\rangle_{\{J\} } \, ,
\end{equation}
where
\begin{equation}\label{5}
{\rm Sp}_{\sigma} (\dots) = \prod_i \sum_{\sigma_i=\pm 1} (\dots) \,
,
\end{equation}
$\beta=T^{-1}$ is the inverse temperature and the averaging over links reads, cf. Equation (\ref{3}):
\begin{equation}\label{6}
\langle (\dots) \rangle_{\{J\} } = \prod_{i<j}  \Big [
(\dots)_{J_{ij}=J} p_{ij} + (\dots)_{J_{ij}=0} (1-p_{ij}) \Big ]  \,
.
\end{equation}

In turn, obtained after averaging over random linking, the partition function ${\cal Z}_N(\{k\})$ depends on the particular choice of  random variable (label) sequence $\{k\}$. Recall that this sequence was taken as a fixed one, i.e.,~quenched. 
Therefore, the observable free energy $F_N$ is to be obtained by averaging
the sequence-dependent free energies $F_N(\{k\})$ as:
\begin{equation}\label{7}
F_N=\langle F_N(\{k\}) \rangle_{\{k\} } = -T \prod_{i} \sum_{k_i} p(k_i)  \ln {\cal Z}_N(\{k\})\, .
\end{equation}
It is worth mentioning here another prominent feature of the annealed network: as we will
explicitly show below, the partition function ${\cal Z}_N(\{k\})$ is self-averaging, i.e.,~it does not
depend on a particular choice of  $\{k\}$:  ${\cal Z}_N(\{k\})\equiv{\cal Z}_N$. This leads to an obvious
relation:
\begin{equation}\label{8}
F_N= -T \prod_{i} \sum_{k_i} p(k_i)  \ln {\cal Z}_N = -T \ln {\cal Z}_N\, ,
\end{equation}
which means that the free energy is a self-averaged quantity too and avoids averaging of the logarithm of partition
function, facilitating calculations on annealed networks.

\subsection{Ising Model with Random Spin Length on an Annealed Network}\label{II.2}

The model we consider in this study \cite{Krasnytska20} relaxes the restriction on the
fixed spin length in the Hamiltonian  (\ref{2a}). Similar to the
Ising model, we preserve the binary character of spin variables
keeping global $Z_2$ symmetry of the whole system, however, we
allow each spin to change its absolute value in a continuous and
random fashion. Namely, we endow the spins $\sigma_i$ with
'strengths' which vary from site to site through a  random
variable $|\sigma_i|\equiv \mathcal{S}_i$. The Hamiltonian of the
model reads:
\begin{equation}\label{9}
{\cal H} = -\frac{1}{2}\sum_{i\neq j} J_{ij} S_i S_j
- H \sum_i S_i \, , \hspace{1cm} S_i=\pm
\mathcal{S}_i\, ,
\end{equation}
where all notations are as in Equation (\ref{2a}) and $\mathcal{S}_i$ are
independent identically distributed (i.i.d.)  random variables with
a given distribution function $q(\mathcal{S})$ each. 
The  Hamiltonian
(\ref{9}) can be equivalently rewritten in terms of usual Ising
spins of unit length, choosing variables $S_i=\sigma_i
\mathcal{S}_i$:
\begin{equation}\label{9'}
{\cal H} = - \frac{1}{2}\sum_{i\neq j} J_{ij} \mathcal{S}_i
\mathcal{S}_j\sigma_i \sigma_j - H \sum_i \mathcal{S}_i\sigma_i \, ,
\hspace{1cm} \sigma_i=\pm 1\, ,
\end{equation}

We consider the case when the sequence  $\{\mathcal{S}
\}=(\mathcal{S}_{\rm min}, ..., \mathcal{S}_{\rm max})$ is maximally
random under the constraint that their distribution is a given one.
For the fixed sequence of random variables $\{k\}$ (that define
network linkage) and $\{ \mathcal{S}\}$ (that define local spin
strength), the partition function is obtained by averaging with
respect to random annealed links $\{J\}$, cf. Equation (\ref{4}):
\begin{equation}\label{10}
{\cal Z}_N(\{k\},\{ \mathcal{S}\}) = \langle {\rm Sp}_{{\sigma}}
e^{-\beta {\cal H}} \rangle_{\{J\} } \, ,
\end{equation}
with the trace  defined in (\ref{5}).

 Generally speaking, after the trace over
spins has been taken, the partition function also remains dependent
on the (randomly distributed) spin strengths $\{ \mathcal{S}\}$, as
explicitly denoted in Equation (\ref{10}). However, in the next subsection,
we show that in the case of annealed networks, the partition function ${\cal
Z}_N(\{k\},\{  \mathcal{S} \})$ is a self-averaging quantity both with
respect to random variables $k$ and $\mathcal{S}$ (${\cal Z}_N(\{k\},\{
\mathcal{S} \}) ={\cal Z}_N$). Therefore, for the free energy, similar
to (\ref{8}), one obtains:
\begin{equation}\label{12}
F_N= -T \prod_{i} \sum_{k_i} p(k_i) \sum_{\mathcal{S}_i}
q(\mathcal{S}_i) \ln {\cal Z}_N(\{k\},\{ \mathcal{S}\}) = -T
\ln {\cal Z}_N\, .
\end{equation}
Our task now is to proceed in deriving the partition function
of the Ising model with varying spin length $\mathcal{S}$ on an
annealed scale-free network when distributions of
the random variables $q(\mathcal{S})$, $p(k)$ follow
power-law behavior (\ref{2.1}), (\ref{2.2}).
In the course of derivation, we arrive at the conclusion
about its self-averaging properties.

\subsection{Self-Averaging}\label{II.3}

Substituting into (\ref{9'}) the adjacency
matrix (\ref{3}) and averaging over spin configurations, we obtain:
\begin{equation}\label{11}
{\cal Z}_N(\{k\},\{\mathcal{S}\})={\rm Sp}_{\sigma} \Big(e^{\beta
H\sum_i \mathcal{S}_i\sigma_i} \prod_{i<j}(p_{ij}e^{\frac{\beta
J}{2}\sum_{i\neq j}\mathcal{S}_i
\mathcal{S}_j\sigma_i\sigma_j}+1-p_{ij}) \Big).
  \end{equation}
Taking into account that the spin product in (\ref{11}) can attain
only two values ($\sigma_i \sigma_j=\pm 1$), we can make use of the
equality
\begin{equation}\label{equality}
f(K\varepsilon)=\frac{1}{2}[f(K)+f(-K)]+\frac{\varepsilon}{2}[f(K)-f(-K)],
\hspace{1cm} \varepsilon = \pm 1,
  \end{equation}
to obtain the partition function (\ref{11}) in case $\varepsilon\equiv
\sigma_i\sigma_j$, $K\equiv\beta J \mathcal{S}_i \mathcal{S}_j$:
\begin{equation}\label{21} 
{\cal Z}_N(\{k\},\{\mathcal{S}\})={\rm Sp}_{S} \Big(e^{\beta H\sum_i
\mathcal{S}_i\sigma_i} \prod_{i<j} \Big([\cosh (\beta J \mathcal{S}_i \mathcal{S}_j)+
\sigma_i\sigma_j\sinh (\beta J
\mathcal{S}_i\mathcal{S}_j)-1]p_{ij}+1\Big).
  \end{equation}
Simplifying the expression for the partition function, one arrives at:
\begin{equation}\label{22}
{\cal Z}_N(\{k\},\{\mathcal{S}\})={\rm Sp}_{\sigma} \Big(e^{\beta
H\sum_i \mathcal{S}_i\sigma_i} \prod_{i<j}
e^{\ln(a_{ij}+b_{ij}\sigma_i\sigma_j)}\Big)
  \end{equation}
with
\begin{equation}\label{23}
a_{ij}=1-p_{ij}+p_{ij}\cosh(\beta J\mathcal{S}_i\mathcal{S}_j),
\hspace{1cm} b_{ij}=p_{ij}\sinh(\beta J\mathcal{S}_i\mathcal{S}_j).
  \end{equation}
  Making use of the equality
(\ref{equality}) to represent $\ln(a_{ij}+b_{ij}\sigma_i\sigma_j)$ in (\ref{22}), we
obtain for the partition function:
\begin{equation}\label{24}
{\cal Z}_N(\{k\},\{\mathcal{S}\})=\prod_{i<j}c_{ij}{\rm Sp}_{\sigma}
\Big(
 e^{\frac{1}{2}\sum_{i\neq j}d_{ij} \sigma_i\sigma_j+\beta H\sum_i
\mathcal{S}_i\sigma_i}\Big),
  \end{equation}
  with
\begin{equation}\label{25}
c_{ij}=\sqrt{a_{ij}^2-b_{ij}^2}, \hspace{1cm}d_{ij}=\ln
\frac{a_{ij}+b_{ij}}{a_{ij}-b_{ij}}.
  \end{equation}
The latter coefficients implicitly depend on $p_{ij}$ via
(\ref{23}). Substituting these dependencies into (\ref{25}), one
obtains:
\begin{equation}\label{26}
c_{ij}=\sqrt{1-2p_{ij}+2p_{ij}^2+2(1-p_{ij})\cosh(\beta
J\mathcal{S}_i\mathcal{S}_j)},
  \end{equation}
\begin{equation}\label{27}
d_{ij}=\ln \frac{1-p_{ij}+p_{ij}e^{\beta J\mathcal{S}_i
\mathcal{S}_j}}{1-p_{ij}+p_{ij}e^{-\beta J \mathcal{S}_i
\mathcal{S}_j}}.
  \end{equation}

\textls[-15]{Substituting
$p_{ij}$ into the expression for the partition function (\ref{24}) and
evaluating $d_{ij}$ (\ref{27}) in the thermodynamic limit $N\to \infty$
(i.e.,~in the limit
of small $p_{ij}$ ),}
\begin{equation}\label{III.1a}
 d_{ij}=\ln \frac{1-p_{ij}+p_{ij}e^{\beta
J\mathcal{S}_i \mathcal{S}_j}}{1-p_{ij}+p_{ij}e^{-\beta
J\mathcal{S}_i\mathcal{S}_j}}\simeq p_{ij}\beta
J\mathcal{S}_i\mathcal{S}_j,
\end{equation}
we get:
\begin{equation}\label{III.1}
{\cal Z}_N(\{k\},\{\mathcal{S}\})={\rm Sp}_{\sigma}  \exp \Big (\beta
J\sum_{i<j}\frac{k_ik_j\mathcal{S}_i\mathcal{S}_j\sigma_i\sigma_j}{N\langle
k \rangle}+\beta H\sum_i \mathcal{S}_i \sigma_i \Big )\,.
\end{equation}

Now the interaction term in (\ref{III.1})
attains a separable form and one can apply Stratonovich--Hubbard
transformation to  take the trace over spins $\sigma_i$ exactly and to obtain
the following expression for the partition function:
\begin{equation}\label{III.2}
{\cal Z}_N(\{k\},\{\mathcal{S}\})= \int_{-\infty}^{+\infty} \exp\Big
(\frac{-N\langle k \rangle T x^2}{2J}+\sum_{i}\ln \cosh [\mathcal{S}_i( xk_i+H/T)]\Big)dx \, .
\end{equation}
In this and all other partition function integral representations, we omit the prefactors that are irrelevant for our analysis.
As long as the functional dependence on the random variables $\mathcal{S}_i$, $k_i$
in (\ref{III.2}) is of the unary type, it is convenient to pass from sums over nodes $i$ to
sums over the random variables $k_i$, $\mathcal{S}_i$  with a given
distribution function $p(k)$, $q(\mathcal{S})$. Considering the random variables to
be continuous, one arrives at:
\begin{equation}\label{III.3}
\sum_i f(k_i{,} \mathcal{S}_i)=N\sum_{k_{\rm min}}^{k_{\rm
max}}\sum_{\mathcal{S}_{\rm min}}^{\mathcal{S}_{\rm
max}}p(k)q(\mathcal{S})f(k, \mathcal{S})\, =
N\int_{k_{\rm min}}^{k_{\rm max}}\int_{\mathcal{S}_{\rm
min}}^{\mathcal{S}_{\rm max}}p(k)q(\mathcal{S})f(k, \mathcal{S})dk
d\mathcal{S} \, .
\end{equation}
For an infinite system, we  put $k_{\rm max}=\mathcal{S}_{\rm
max}\to \infty$ and, without a loss of generality, we choose the lower
bonds equal to $k_{\rm min}=\mathcal{S}_{\rm min} = 2$ and $J=1$. Note, that the peculiarities of the critical behavior we are interested
 in are caused by the behavior at $k_{\rm max},\mathcal{S}_{\rm
max}\to \infty$. Although it is more natural to choose the lower integration bond equal to unity,
scale-free networks with $k_{\rm min}=1$ do not possess a spanning cluster for
$\lambda>\lambda_c$ (with $\lambda_c=3.48$ for discrete node degree distribution
and $\lambda_c=4$  for the continuous one) \cite{Cohen02,Aiello00,Krasnytska13}. 
We avoid this restriction by choosing $k_{\rm min}=2$. To have expressions symmetric in $k,\mathcal{S}$, we choose $\mathcal{S}_{\rm min}=2$ too. 
Now it is straightforward to see that the partition function  ${\cal
Z}_N(\{\mathcal{S}\},\{k\})$ does not depend on random variables $k$
and $\mathcal{S}$ and is {\em self-averaging}:

\end{paracol}
\nointerlineskip
\begin{equation}\label{III.3b}
{\cal Z}_N(\{k\},\{\mathcal{S}\}) \equiv {\cal Z}_N =
\int_{-\infty}^{+\infty} \exp\Big (\frac{-N\langle k \rangle T
x^2}{2}+ N\int_{2}^{\infty}\int_{2}^{\infty}p(k)q(\mathcal{S})\ln \cosh[\mathcal{S}(
kx+H/T)]dkd\mathcal{S}\Big)dx\, .
\end{equation}
\begin{paracol}{2}
\switchcolumn

As one can see from Equation (\ref{III.3b}), the self-averaging property is quite general and 
concerns any form of distributions $p(k)$, $q(\mathcal{S})$. Below, we use this expression
to analyze thermodynamics in the case when these distributions attain power-law forms 
(\ref{2.1}), (\ref{2.2}). 
 
\section{Thermodynamic Functions}\label{IV}

It is convenient to pass in Equation (\ref{III.3b}) to integration over positive values of  $x$  and to present
the partition function   
as
\begin{eqnarray}\nonumber
{\cal Z}_N&=& \int_{0}^{+\infty} e^{\frac{-\langle k \rangle
x^2T}{2N}} \Big [ \exp \Big (N\int_2^{\infty}\int_2^{\infty}p(k)q(\mathcal{S})\ln \cosh (\frac{\mathcal{S}
kx}{N}+\mathcal{S}H/T)dkdL\Big ) + \\&& \label{III.2a} \exp \Big (N\int_2^{\infty}\int_2^{\infty}p(k)q(\mathcal{S})\ln \cosh (\frac{-\mathcal{S}
kx}{N}+\mathcal{S}H/T)dkd\mathcal{S} \Big ) \Big ]dx\, .
\end{eqnarray}

Being interested in the leading asymptotics of the partition function
at $N\to \infty$ and  keeping the
first leading term in $H$, we present
the expression (\ref{III.2a}) in the following form:
\begin{equation} \label{III.5}
{ \cal Z}_N = \int_{0}^{+\infty}  e^{\frac{-\langle k \rangle
x^2T}{2N}} \Big
[\exp(I^{+}_{\lambda,\mu}(x))+\exp(I^{-}_{\lambda,\mu}(x)) \Big ]
\,dx\, ,
\end{equation}
with
\begin{equation}\label{III.6}
I^{\pm}_{\lambda,\mu}(x)=N \Big [c_\lambda c_\mu \Big(\frac{x}
{N}\Big)^{\frac{\lambda+\mu-2}{2}}I_{\lambda,\mu}(\varepsilon)
\pm\frac{\langle \mathcal{S}^2\rangle \langle k\rangle}{TN}xH\Big]
\end{equation}
 where
\begin{equation}\label{III.6a}
I_{\lambda,\mu}(\varepsilon)=
\int_{\varepsilon}^{\infty}\int_{\varepsilon}^{\infty} \frac{\ln
\cosh( k\mathcal{S})}{k^\lambda \mathcal{S}^\mu} \, d\mathcal{S}dk \,
\end{equation} 
and we have substituted distributions  $q(\mathcal{S})$, $p(k)$ in power-law forms (\ref{2.1}) and
(\ref{2.2}).
The lower integration bound $\varepsilon=2 \sqrt{\frac{x}{N}}$ tends to zero, when $N\rightarrow \infty$. 
The asymptotic expansions of the integral (\ref{III.6a}) at small $\varepsilon$ (large $N$) are 
evaluated in the Appendix.
Substituting these expansions at different values of parameters $\lambda$, $\mu$ into Equation (\ref{III.5}),
we arrive at corresponding expressions for the partition function that is evaluated
at large $N$ by the steepest descent method. The final expression for the partition function reads:
\begin{equation}\label{IV.1}
{ \cal Z}_N = \int_{0}^{+\infty} e^{-N\Phi_{\mu,\lambda}(x)} \,dx\, ,
\end{equation}
{where} 
\begin{equation}\label{IV.2}
\Phi_{\mu,\lambda}(x) = \frac{\langle k \rangle x^2T}{2}-c_\mu
c_\lambda
x^{\frac{\lambda+\mu-2}{2}}I_{\lambda,\mu}(\sqrt{x})
-\frac{\langle \mathcal{S}^2\rangle \langle k\rangle}{T}xH \, 
\end{equation}
and the linear term in $H$ originates from the large $N$ asymptotics of the hyperbolic cosine in Equations (\ref{III.5}) and (\ref{III.6}). 

Now it is straightforward to write for the Helmholtz free energy 
$F_N(T,H)$ per node:
\begin{equation}\label{IV.2a}
f(T,H)=\lim_{N\to \infty} F_N(H,T)/N= - T \lim_{N\to \infty} \ln { \cal Z}_N/N = T \Phi_{\mu,\lambda}(m) 
\end{equation}
with $m$ being the coordinate of function $\Phi_{\mu,\lambda}(x)$ minimum:
\begin{equation}
\frac{{\rm d}\, \Phi_{\mu,\lambda}(x)}{{\rm d}\, x} |_{x=m} =0,
\hspace{3em} \frac{{\rm d}^2\, \Phi_{\mu,\lambda}(x)}{{\rm d}\, x^2} |_{x=m} > 0\, .
  \end{equation}
The resulting free energy is symmetric upon an interchange of
indices $\mu\leftrightarrow \lambda$. Therefore, below, we give the
corresponding expressions for two cases:  $\mu >\lambda$ and $\mu=\lambda$. 
For the first case, $\mu >\lambda$, an {asymptotic} of the free energy 
at small $m$ is governed by the lower value of the exponents, i.e.,~by $\lambda$. 
Keeping the leading terms, we arrive at:  
\begin{adjustwidth}{-4.6cm}{0cm}
\begin{eqnarray}\label{A10''}
\Phi_{\mu,\lambda}(m)+\frac{\langle \mathcal{S}^2\rangle \langle k\rangle}{T}mH \simeq  \left\{
\begin{array}{lll}
2 < \lambda <3: &&  -\frac{c_\mu c_\lambda i_\lambda}{\mu-\lambda} m^{\lambda-1} +\frac{\langle k\rangle T}{2}m^2,  \\
\lambda =3: &&  \frac{c_\mu c_3}{2(\mu-3)} m^2\ln \frac{1}{m}+
c_\mu c_3m^2(\frac{i_3}{3-\mu}+\frac{1}{2(\mu-3)^2})+\frac{\langle k\rangle T}{2} m^2,  \\
3 < \lambda < 5: && \frac{\langle k\rangle}{2}(T-T_0)m^2-\frac{c_\mu c_\lambda i_\lambda}{\mu-\lambda}
m^{\lambda-1},  \\
\lambda = 5: &&
\frac{\langle k\rangle}{2}(T-T_0)m^2-\frac{c_5c_\mu}{12(\mu-5)}m^4\ln
\frac{1}{m} + c_5c_\mu (\frac{1}{12(\mu-5)^2}-\frac{i_5}{\mu-5})m^4,  \\
\lambda>5: && \frac{\langle k\rangle }{2}(T-T_0)m^2+ \frac{c_\mu c_\lambda}{12(\lambda-5)(\mu-5)}m^4,
\end{array}
\right.
\end{eqnarray}
\end{adjustwidth}
with
\begin{equation}\label{T0}
T_0= {\frac{c_\mu c_\lambda}{\langle k\rangle (\lambda-3)(\mu-3)}}=\frac{\langle k^2\rangle \langle  \mathcal{S}^2\rangle}{2^{3-\mu}2^{3-\lambda}\langle k\rangle}\, ,
\end{equation}
where $\langle \mathcal{S}^2\rangle=\int_2^\infty  \mathcal{S}^2 q(\mathcal{S}) d\mathcal{S}$, 
$\langle k^2\rangle=\int_2^\infty  k^2 p(k) dk $, the distribution functions 
$q(\mathcal{S})$, $p(k)$ are given by Equations (\ref{2.1}) and (\ref{2.2}),
and we have taken into account that 
$\mathcal{S}_{\rm min}=k_{\rm min}=2$ (see explanation below Eq. (\ref{III.3})).  
The coefficients $i_\mu$  are listed in the Appendix and $c_\mu$, 
$c_\lambda$ are normalizing factors of the distribution functions (\ref{2.1}), (\ref{2.2}).

For the case $\lambda=\mu$, the leading behavior at small $m$ reads:
\begin{adjustwidth}{-4.6cm}{0cm}
\begin{eqnarray}\label{A10'''}
\Phi_{\mu,\mu}(m)+\frac{\langle \mathcal{S}^2\rangle \langle k\rangle}{T}mH \simeq \left\{
\begin{array}{lll}
2 < \mu <3: && -c_\mu^2 i_\mu  m^{\mu-1} \ln \frac{1}{m}-c_3^2 i_{\mu,\mu}m^{\mu-1}+\frac{\langle k\rangle T}{2}m^2,  \\
\mu =3: &&-i_3c_3^2 m^2 \ln \frac{1}{m} + [\frac{\langle k\rangle T}{2}-c_3^2 i_{3,3}]m^2,  \\
3 < \mu < 5: && \frac{\langle k\rangle}{2}(T-T_0)m^2-c_\mu^2i_\mu
m^{\mu-1} \ln \frac{1}{m}\, ,  \\
\mu = 5: &&
\frac{\langle k\rangle}{2}(T-T_0)m^2-\frac{c_5^2}{24}m^4(\ln \frac{1}{m})^2-i_5 c_5^2 m^4\ln \frac{1}{m}\, ,  \\
\mu>5: && \frac{\langle k\rangle}{2}(T-T_0)m^2+
\frac{c_\mu^2}{12(\mu-5)^2}m^4\, ,
\end{array}
\right.
\end{eqnarray}
\end{adjustwidth}
with the notations explained above. The signs of the coefficients $i_{\mu,\lambda}$ do not matter in our analysis.

The estimates obtained above for the free energy asymptotics (\ref{A10''}), (\ref{A10'''}) give one access to the thermodynamic properties of the system
of interest. As we will see below, parameters
$\mu$ and $\lambda$ play a crucial role in governing the onset of ordering and define the universality class of the generalized Ising model on a 
scale-free network. Before proceeding in analyzing these expressions,
it is instructive to recall the main peculiarities of the critical
behavior of two models, where each of these parameters has been considered separately: these are the Ising model on a scale-free network with a node-degree distribution (\ref{2.2}) \cite{Leone02,Goltsev03,Palchykov11} and the generalized Ising model with a power-law
spin strength distribution (\ref{2.1}) on a complete graph \cite{Krasnytska20}.
As  is well established by now, the Ising model on a scale-free network  remains ordered at any finite temperature at
low values of the node-degree distribution exponent $2<\lambda\leq3$. 
The order parameter 
decays with temperature as a power law  
$m\sim T^{1/(\lambda-3)}$ at $2<\mu < 3$. The decay is exponential
for $\lambda=3$: $m \sim e^{-bT}$. With a further increase in $\lambda$,  
 a second order phase transition occurs for $\lambda>3$ at finite $T=T_0$ and 
$H=0$: $m=0$ at the high-temperature phase, whereas 
the order parameter emerges as $m\sim \tau^{1/(\lambda-3)}$ in the vicinity 
of the  transition point at $H=0$ with $\tau=|T-T_0|/T_0$. The power-law temperature
behavior of the order parameter attains its usual mean-field value only when
$\lambda$ exceeds five: $m\sim \tau^{1/2}$, $\lambda>5$. Logarithmic correction
to scaling appears at marginal $\lambda=5$: $m\sim \tau^{1/2}|\ln \tau|^{-1/2}$.
The  phase diagram  described above is sketched in Figure~\ref{fig2}a. A similar
picture is observed when one analyzes the generalized Ising model with a power-law
spin strength distribution  on a complete graph, i.e.,~when, in the spirit
of the Kac model \cite{Kac1,Kac2,Kac3,Kac4,Kac5,Kac6,Kac7}, each graph node
is connected to all other nodes. As has been demonstrated in Ref.  \cite{Krasnytska20},
the role of the global parameter is played in this case by the spin strength distribution
exponent $\mu$. In turn, we summarize the behavior of the order parameter $m$ for different values
of $\mu$ in Figure~\ref{fig2}c.

 \begin{figure}[H]
{\includegraphics[width=13cm]{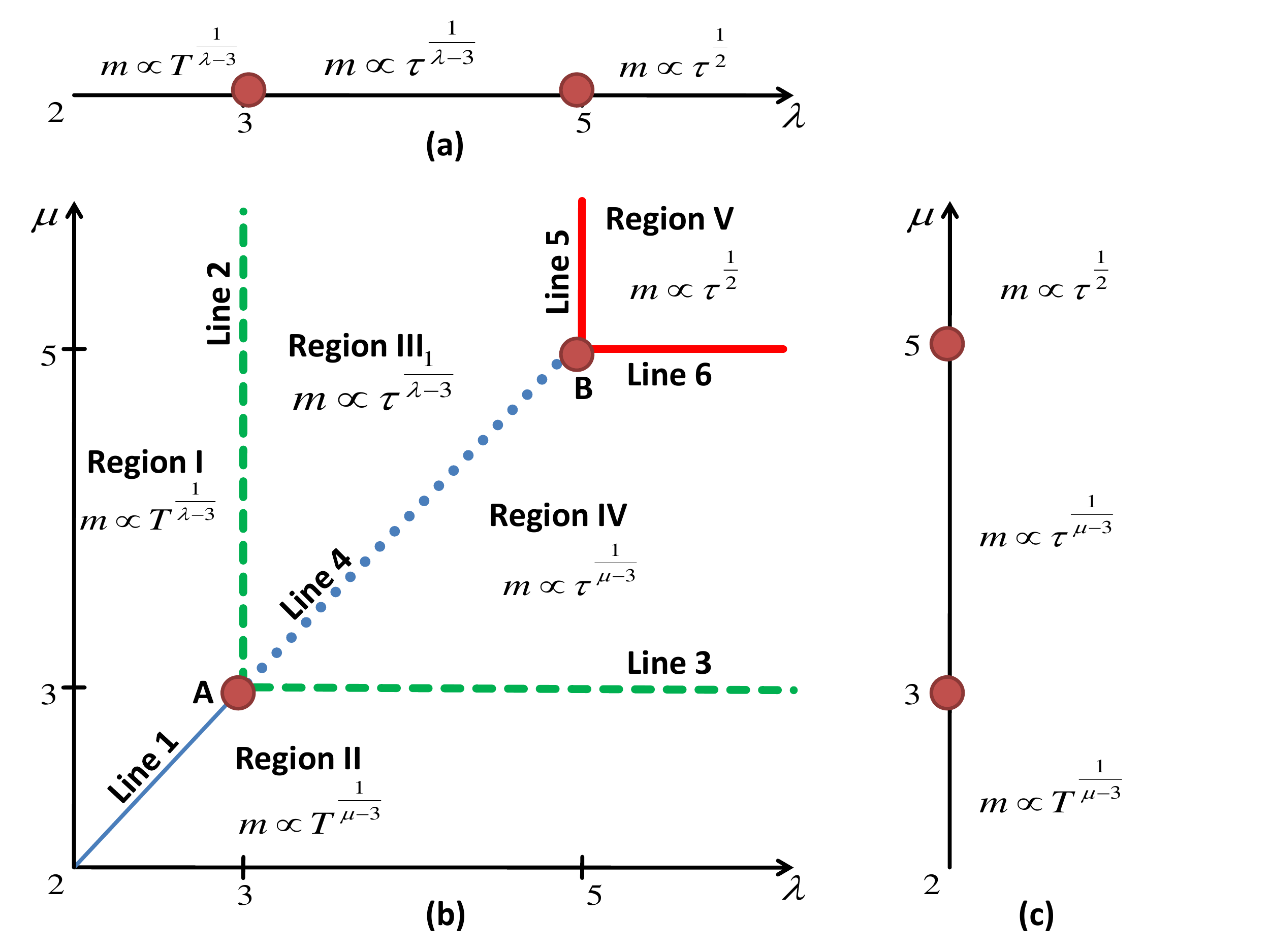}}
\caption{Phase diagram of the generalized Ising model
with power-law distributed spin strength 
on a scale-free network (\textbf{b}) is compared with those
for the Ising model on a scale-free network (\textbf{a}) and generalized
Ising model with power-law distributed spin strength on a 
complete graph (\textbf{c}). Asymptotics of the order parameter
in different regions of $\mu$, $\lambda$ are shown explicitly.
Corresponding asymptotics at marginal values of $\mu$, $\lambda$
(lines and points in the plot) are summarized in Table \ref{tab1}.
  \label{fig2}}
\end{figure}

Now, with the free energy asymptotics for the generalized Ising model 
on a scale-free network (\ref{A10''}), (\ref{A10'''}) at hand, we are in a position to analyze the interplay of two parameters: the first one governing individual
spin strength ($\mu$) and the second one governing its connectivity
($\lambda$), on the emergent critical behavior.
Temperature behavior of the order parameter and  the
phase diagram that originate from this analysis are shown in
 Table \ref{tab1} and in Figure~\ref{fig2}b.
The
behavior is controlled by the parameter ($\lambda $ or $\mu$) with
the smaller value. When at least one of the parameters ($\lambda $
or $\mu$) is less than three, the system remains
ordered at any finite temperature and the order parameter decays as a
power-law function  of $T$:
\begin{eqnarray}\label{23ml}
m \simeq \left\{
\begin{array}{lll}
2 < (\mu,\lambda) <3;\, \mu>\lambda: && T^{\frac{1}{\lambda-3}}, \\ 
2 < (\mu,\lambda) <3;\, \mu=\lambda: && T^\frac{1}{\lambda-3}, \\ 
2 < (\mu,\lambda) <3;\, \mu<\lambda: && T^{\frac{1}{\mu-3}}. \\ 
\end{array}
\right.
\end{eqnarray}
When either $\lambda $ or $\mu$ equals three, and the
other one is larger than three, $m$ decays exponentially. 
A second
order phase transition occurs when both $\lambda,\mu>3$. Depending on
the values of $\lambda,\mu$, the order parameter is characterized
by different asymptotics.  In the
region $3<\mu<5$ ($\mu<\lambda$), the critical exponents are
$\mu$ dependent, and in region $3<\lambda<5$ ($\mu>\lambda$), they are
$\lambda$ dependent and logarithmic corrections appear in these regions
at $\lambda=\mu$:
\begin{eqnarray}\label{35ml}
m \simeq \left\{
\begin{array}{lll}
3 < (\mu,\lambda) <5;\, \mu>\lambda: && \tau^{\frac{1}{\lambda-3}}, \\ 
3 < (\mu, \lambda) <5;\, \mu=\lambda: && (\tau |\ln \tau|^{-1})^{\frac{1}{\lambda-3}}, \\ 
3 < (\mu, \lambda) <5;\, \mu<\lambda: && \tau^{\frac{1}{\mu-3}}. \\ 
\end{array}
\right.
\end{eqnarray} 
Logarithmic corrections to scaling, however, of different values, also appear when 
$\lambda=5$ or $\mu=5$. We discuss these corrections in more detail later.

\begin{specialtable}[H]

\caption{\label{tab1} Temperature behavior of the order  parameter $m$ at different
values of $\mu$ and $\lambda$. The asymptotic is governed by the smaller parameter from the 
pair ($\mu,\lambda$).}
\setlength{\cellWidtha}{\columnwidth/6-2\tabcolsep+0.0in}
\setlength{\cellWidthb}{\columnwidth/6-2\tabcolsep+0.0in}
\setlength{\cellWidthc}{\columnwidth/6-2\tabcolsep-0.0in}
\setlength{\cellWidthd}{\columnwidth/6-2\tabcolsep-0.0in}
\setlength{\cellWidthe}{\columnwidth/6-2\tabcolsep-0.0in}
\setlength{\cellWidthf}{\columnwidth/6-2\tabcolsep-0.0in}
\scalebox{1}[1]{\begin{tabularx}{\columnwidth}{>{\PreserveBackslash\centering}m{\cellWidtha}>{\PreserveBackslash\centering}m{\cellWidthb}>{\PreserveBackslash\centering}m{\cellWidthc}>{\PreserveBackslash\centering}m{\cellWidthd}>{\PreserveBackslash\centering}m{\cellWidthe}>{\PreserveBackslash\centering}m{\cellWidthf}}
\toprule
   & \boldmath{$2<\lambda<3$} & \boldmath{$\lambda=3$} & \boldmath{$3<\lambda<5$} & \boldmath{$\lambda=5$} & \boldmath{$\lambda>5$} \\
  \midrule
  $2<\mu<3$ & Equation (\ref{23ml}) & $ T^{\frac{1}{\mu-3}}$ & $ T^{\frac{1}{\mu-3}}$ &
  $ T^{\frac{1}{\mu-3}}$ & $ T^{\frac{1}{\mu-3}}$ \\
 $\mu=3$ & $ T^{\frac{1}{\lambda-3}}$ & $ e^{-bT}$ & $ e^{-bT}$ & $ e^{-bT}$ & $ e^{-bT}$ \\
  $3<\mu<5$ & $ T^{\frac{1}{\lambda-3}}$ & $ e^{-bT}$ & Equation (\ref{35ml}) & $ \tau^{\frac{1}{\mu-3}}$ & $ \tau^{\frac{1}{\mu-3}}$ \\
  $\mu=5$ & $ T^{\frac{1}{\lambda-3}}$ & $ e^{-bT}$ & $ \tau^{\frac{1}{\lambda-3}}$ & $ \tau^{\frac{1}{2}}|\ln \tau|^{-1}$ & $ \tau^{\frac{1}{2}}|\ln \tau|^{-\frac{1}{2}}$ \\
  $\mu>5$ & $ T^{\frac{1}{\lambda-3}}$ & $ e^{-bT}$ & $ \tau^{\frac{1}{\lambda-3}}$ & $\tau^{\frac{1}{2}}|\ln \tau|^{-\frac{1}{2}}$ & $ \tau^{\frac{1}{2}}$ \\
  \bottomrule 
\end{tabularx}}

\end{specialtable}

The phase diagram in Figure~\ref{fig2}b visualizes the behavior discussed above. 
There, we show different regions in the $\lambda-\mu$ plane that are characterized by
different critical behaviors. The last is governed by the distribution with a 'fatter'
tail (smaller value from the pair $\lambda,\mu$). It is instructive to compare
this diagram with those of Figure \ref{fig2}a,c. Indeed, when one of the exponents
in Figure~\ref{fig2}b is  larger than five (very fast decay of one of the distributions 
(\ref{2.1}) or (\ref{2.2})), the resulting diagram does not depend on this exponent
any more. One may speak about degeneracy of the critical behavior with respect to this
exponent and about reduction of the phase diagram Figure~\ref{fig2}b to one of its 
corresponding counterparts, as shown in Figure~\ref{fig2}a,c. Interesting new phenomena emerge along 
the lines of the diagram in Figure~\ref{fig2}b, that separate regions with different {asymptotics}
of the order parameter. Usually, changes in the power law {asymptotics} of thermodynamic observables
are accompanied by logarithmic correction-to-scaling exponents (see, e.g.,~\cite{Kenna12} and references
therein). For $d$-dimensional lattices, such corrections appear at upper critical dimensions,
and for the scale-free networks they are known to {accompany the} leading asymptotics at $\lambda=5$.
In our analysis, we complete the picture by observing the {\em lines} in the 
$\lambda-\mu$ plane, where such corrections appear. Furthermore, new scaling laws are observed
at the intersection of these lines, as further outlined below.

To proceed with the analysis of critical behavior, we obtain expressions
for the other thermodynamic functions in the vicinity of the second order 
phase transition that occurs for $\mu,\lambda>3$ at $T=T_0$, $H=0$.
In particular, besides the order parameter,  we evaluate the 
leading critical exponents for the isothermal susceptibility $\chi_T$,
specific heat $c_H$, and magnetocaloric {coefficient} $m_T$ (the magnetocaloric coefficient is defined by the mixed 
derivative of the free energy over magnetic field and temperature,
$m_T=-T(\partial m/\partial T)_H$):
\begin{equation}\label{15}
m\sim\tau^\beta,\hspace{1cm} \chi_T \sim\tau^{-\gamma}, \hspace{1cm}
c_H\sim\tau^{-\alpha},\hspace{1cm} m_T \sim \tau^{-\omega}, \hspace{2em} {\rm at} \hspace{2em} H=0\, .
\end{equation}
\begin{equation}\label{16}
m\sim H^{1/\delta}, \hspace{1cm} \chi_T\sim H^{-\gamma_c},
\hspace{1cm} c_H\sim H^{-\alpha_c}, \hspace{1cm} m_T \sim
H^{-\omega_c}, \hspace{1em} {\rm at} \hspace{1em} \tau=0\, .
\end{equation}
We also find the logarithmic terms that appear at marginal values
of $\lambda$, $\mu$ and define the logarithmic correction exponents
for each of the above quantities:
\begin{equation}\label{log}
A \sim \tau^{\Theta} |\ln \tau|^{\hat{\Theta}}\, ,
\hspace{2em} H=0\, . \hspace{1cm} A \sim
H^{\Theta_c} |\ln H|^{\hat{\Theta}_c}\, , \hspace{2em} \tau=0\, ,
\end{equation}
where $A$ is one  of the thermodynamic functions (\ref{16}), $\Theta$ is the critical exponent,
and $\hat{\Theta}$ is a corresponding logarithmic correction exponent. 
Values of the leading critical exponents for thermodynamic functions (\ref{15}) and (\ref{16}) 
are summarized in Table \ref{tab2}. The corresponding logarithmic corrections to
scaling exponents are collected in Table \ref{tab3}.

\begin{specialtable}[H]

\caption{Critical indices of the generalized model with power-law distributed spin 
strength on an annealed scale-free network in different regions of the
phase diagram Figure~\ref{fig2}b. Line 4:
$3< (\lambda,\mu) <5$, $\lambda=\mu$; region III: $3<\mu<5$, $\mu<\lambda$;
region IV: $3<\lambda<5$, $\lambda<\mu$; region V: $\lambda,
\mu \geq 5$. \label{tab2}}

\setlength{\cellWidtha}{\columnwidth/9-2\tabcolsep+0.8in}
\setlength{\cellWidthb}{\columnwidth/9-2\tabcolsep-0.1in}
\setlength{\cellWidthc}{\columnwidth/9-2\tabcolsep-0.1in}
\setlength{\cellWidthd}{\columnwidth/9-2\tabcolsep-0.1in}
\setlength{\cellWidthe}{\columnwidth/9-2\tabcolsep-0.1in}
\setlength{\cellWidthf}{\columnwidth/9-2\tabcolsep-0.1in}
\setlength{\cellWidthg}{\columnwidth/9-2\tabcolsep-0.1in}
\setlength{\cellWidthh}{\columnwidth/9-2\tabcolsep-0.1in}
\setlength{\cellWidthi}{\columnwidth/9-2\tabcolsep-0.1in}%
\scalebox{1}[1]{\begin{tabularx}{\columnwidth}{>{\PreserveBackslash\centering}m{\cellWidtha}>{\PreserveBackslash\centering}m{\cellWidthb}>{\PreserveBackslash\centering}m{\cellWidthc}>{\PreserveBackslash\centering}m{\cellWidthd}>{\PreserveBackslash\centering}m{\cellWidthe}>{\PreserveBackslash\centering}m{\cellWidthf}>{\PreserveBackslash\centering}m{\cellWidthg}>{\PreserveBackslash\centering}m{\cellWidthh}>{\PreserveBackslash\centering}m{\cellWidthi}}
\toprule
  & \boldmath{$\alpha$} & \boldmath{$\alpha_c$} & \boldmath{$\gamma$} & \boldmath{$\gamma_c$} &
 \boldmath{$\beta$} &\boldmath{$\delta$} &
\boldmath{$\omega$} & \boldmath{$\omega_c $} \\
\midrule
Line 4 ($\mu=\lambda$) & $\frac{\lambda-5}{\lambda-3} $   & $\frac{\lambda-5}{\lambda-2}$ & 1 &   $\frac{\lambda-3}{\lambda-2}$ &  $ \frac{1}{\lambda-3}$ &  $\lambda-2 $ &  $\frac{\lambda-4}{\lambda-3}$ &  $\frac{\lambda-4}{\lambda-2}$  \\

Region III & $\frac{\lambda-5}{\lambda-3} $   & $\frac{\lambda-5}{\lambda-2}$ & 1  &  $\frac{\lambda-3}{\lambda-2}$ &  $ \frac{1}{\lambda-3}$ &  $\lambda-2 $ &  $\frac{\lambda-4}{\lambda-3}$ &  $\frac{\lambda-4}{\lambda-2}$  \\

Region IV & $\frac{\mu-5}{\mu-3} $   & $\frac{\mu-5}{\mu-2}$ &  1 &  $\frac{\mu-3}{\mu-2}$ &  $ \frac{1}{\mu-3}$ &  $\mu-2 $ &  $\frac{\mu-4}{\mu-3}$ &  $\frac{\mu-4}{\mu-2}$  \\

Region V, Lines 5--6, B & 0    & 0 &  1  &  2/3 &   1/2 &  3 &  1/2 &  1/3  \\
\bottomrule
\end{tabularx}
}

\end{specialtable}

Similar to the case of scale-free networks, the logarithmic corrections
to scaling appear at $\lambda = 5$, $\mu>5$, and $\mu=5$, $\lambda>5$,  along Lines 5 and 6 in Figure~\ref{fig2}b.   
The values of the logarithmic correction exponents coincide with those for the usual Ising model on a scale-free network 
\cite{Leone02,Goltsev03,Palchykov09}.  However, two new 
types of logarithmic corrections emerge in the model under consideration: in region $3< (\lambda=\mu) <5$ (line 4 in Figure~\ref{fig2}b ) 
as well as at $\lambda=\mu=5$ (point B). For $\lambda=\mu=5$,  all logarithmic correction exponents are twice as large in comparison 
with those for the Ising model on a scale-free network at $\lambda=5$. In the region $3<(\lambda=\mu)<5$, all logarithmic correction exponents are $\lambda$ dependent. All of them obey the 
scaling relations for logarithmic corrections  \cite{Kenna061,Kenna062,Kenna063}. 

\begin{specialtable}[H]
  \caption{Logarithmic correction exponents of the generalized model with power-law distributed spin 
strength on an annealed scale-free network in
            different regions. Exponents for lines 5-6 coincide with those found previously 
            \cite{Leone02,Goltsev03,Palchykov09}.
            Here, we find two new sets of exponents that govern logarithmic corrections along
            line 4 and in point B. \label{tab3}}

 \setlength{\cellWidtha}{\columnwidth/9-2\tabcolsep+0.8in}
\setlength{\cellWidthb}{\columnwidth/9-2\tabcolsep-0.1in}
\setlength{\cellWidthc}{\columnwidth/9-2\tabcolsep-0.1in}
\setlength{\cellWidthd}{\columnwidth/9-2\tabcolsep-0.1in}
\setlength{\cellWidthe}{\columnwidth/9-2\tabcolsep-0.1in}
\setlength{\cellWidthf}{\columnwidth/9-2\tabcolsep-0.1in}
\setlength{\cellWidthg}{\columnwidth/9-2\tabcolsep-0.1in}
\setlength{\cellWidthh}{\columnwidth/9-2\tabcolsep-0.1in}
\setlength{\cellWidthi}{\columnwidth/9-2\tabcolsep-0.1in}%
\scalebox{1}[1]{\begin{tabularx}{\columnwidth}{>{\PreserveBackslash\centering}m{\cellWidtha}>{\PreserveBackslash\centering}m{\cellWidthb}>{\PreserveBackslash\centering}m{\cellWidthc}>{\PreserveBackslash\centering}m{\cellWidthd}>{\PreserveBackslash\centering}m{\cellWidthe}>{\PreserveBackslash\centering}m{\cellWidthf}>{\PreserveBackslash\centering}m{\cellWidthg}>{\PreserveBackslash\centering}m{\cellWidthh}>{\PreserveBackslash\centering}m{\cellWidthi}}
\toprule  & \boldmath{$\hat{\alpha}$} & \boldmath{$\hat{\alpha_c}$} & \boldmath{$\hat{\gamma}$}
                & \boldmath{$\hat{\gamma_c}$} &
                \boldmath{$\hat{\beta}$} &\boldmath{$\hat{\delta}$} &
                \boldmath{$\hat{\omega}$} & \boldmath{$\hat{\omega_c} $} \\
                \midrule
                Line 4 ($\mu=\lambda$) & $-\frac{3}{\lambda-2} $   & $-\frac{3}{\lambda-2}$ & 0 &   $-\frac{\lambda-3}{2(\lambda-2)}$ &  $ -\frac{1}{\lambda-3}$ &  $ -\frac{1}{\lambda-2}$&  $-\frac{\lambda-4}{\lambda-3}$ &  $-2\frac{\lambda-4}{\lambda-2}$  \\
                Point B & $-2$  & $-2$ &  $0$ &  $-2/3$ &   $-1$ &  $-2/3$ &  $-1$ &  $-4/3$ \\
                Lines 5--6 & $-1$  & $-1$ &  $0$ &  $-1/3$ &   $-1/2$ &  $-1/3$ &  $-1/2$ &  $-2/3$ \\
                \bottomrule
            \end{tabularx}
        }

    \end{specialtable}

\section{Conclusions and Outlook}\label{V}

The effects of structural disorder on the onset of magnetic ordering in regular (lattice) systems is of mainstream interest in the modern theory of phase transitions and critical phenomena \cite{Holovatch04_20_1,Holovatch04_20_2,Holovatch04_20_3,Holovatch04_20_4,Holovatch04_20_5,Holovatch04_20_6}. 
It is well established by now that even a weak dilution by non-magnetic components may lead to crucial changes in the behavior of magnetically ordered systems. 
If such a dilution is implemented in a quenched fashion, changes in the universality class of the Ising model \cite{Folk03} are governed by the Harris criterion \cite{Harris74}. 
Annealed dilution, on the other hand, causes changes in the Ising model critical exponents via Fisher renormalization~\cite{Fisher68,KennaHsu08}. 
Another textbook example of structural disorder is given by frustrations that may be implemented in the lattice Ising model by (quenched) competing ferro- and anti-ferromagnetic interactions and they are known to cause the spin-glass phase \cite{spin_glasses1,spin_glasses2}.

The generalized Ising model we consider here relaxes the usual condition of a fixed spin length (spin strength)
and considers it as a quenched random variable with a given probability distribution. In the particular case where this random variable is 1 with probability $p$ and 0 with probability $1-p$, one arrives at the familiar quenched diluted Ising model. 
In this study we consider, however, another, richer case, whereby the random spin strength obeys a power-law  distribution (\ref{2.1}) governed by the exponent $\mu$. 
The model mimics polydispersity in  magnetic moments of elementary interacting spins. 
Being interested in possible applications of such a model in the broad area of complex system science, we have analyzed its behavior on an annealed scale-free network. 
In doing so, we make use of two advantages: the annealed network approximation leads to  self-averaging properties of thermodynamic functions and the scale-free behavior of the node-degree distribution (\ref{2.2}) allows us to study competition of power laws (\ref{2.1}), (\ref{2.2}) in defining critical behavior.

As appeared in the course of our study, the model under consideration possesses a number of interesting unexpected features.
Some of them are summarized in Figure~\ref{fig2}b and Tables \ref{tab1}--\ref{tab3}.
The phase diagram of Figure~\ref{fig2}b is accompanied by two others, Figure \ref{fig2}a,c, that correspond to the
usual Iisng model on a scale-free network (a) and to the generalized Ising model with the
power-law distributed spin strength in the complete graph (c). As one can see from this sketch,
the diagram is symmetric under $\mu\leftrightarrow \lambda$ interchange. This means that both factors 
(i.e.,~node connectivity and individual spin strength) influence criticality in a similar fashion. 
Moreover,  the corresponding asymptotics are governed by the smaller  of the pair of parameters ($\mu,\lambda$): the 'fatter' tail of the distribution function wins the competition in defining universality class!
For very low values $2< (\mu, \lambda) \leq3$, the system remains ordered at any finite temperature. 
In turn, the second order phase transition regime ($\mu, \lambda>3$)  is characterized by three different sets of critical
exponents (see Table \ref{tab2}).

Peculiar phenomena emerge in the regions with $\mu=\lambda$, where the changes in critical exponent $\mu$ or $\lambda$ dependencies occur. 
As one observes from Table \ref{tab1}, such changes are accompanied by an emergence of logarithmic corrections in the form of Equation (\ref{log}). 
The values of the logarithmic correction exponents are summarized in Table \ref{tab3}. 
It is instructive to compare this  phenomenon with what happens to the critical behavior in $d$-dimensional
Euclidean space. 
There, a special role is played by a concept of an upper critical dimension $d_u$. By definition, this is the space dimension above which the universality class is trivially defined by the
mean-field behavior~\cite{KennaBerche12}. 
A special type of logarithmic corrections to scaling appears at the upper critical dimension (see \cite{Kenna12}). 
For the scale-free networks, the logarithmic corrections were known to appear at $\lambda=5$, where leading
exponents attain their mean-field values \cite{Leone02,Goltsev03,Palchykov09}. 
Similar corrections also emerge for the generalized Ising model with the
power-law distributed spin strength on the complete graph at $\mu=5$ \cite{Krasnytska20}. 
For the model considered here, these corrections (observed before at single
points in Figure~\ref{fig2}a,c) are now observed throughout along lines 5, 6 in
Figure~\ref{fig2}b. 
The crossing point of these lines,  point B in Figure~\ref{fig2},
is characterized by a new values of logarithmic corrections. Moreover, 
another new set of logarithmic corrections appears
at $3< \mu= \lambda < 5$.

We are deeply indebted to Thomas Ising for conveying to us many insights into Ernst and his story. We are grateful to Sigismund Kobe for further historical insights. We also thank Reinhard Folk for our common work on historical detail of the Ising model and its development over the past century.
This work was supported in part by the  National Research Foundation of Ukraine, project 2020.01/0338 (M.K.)
and by the National Academy of Sciences of Ukraine, project KPKBK6541230 (Yu.H).

\vspace{6pt} 

\authorcontributions{Conceptualization, Yu.H.;
Methodology, M.K. and Yu.H. and B.B. and R.K.; 
Investigation, M.K. and Yu.H.; Visualization, M.K.; 
Validation, B.B. and R.K.; Writing—original draft preparation, M.K. and Yu.H.;
Writing—review and editing, R.K. and B.B. and M.K. and Yu.H.}

\funding{ This work was supported in part by the  National Research Foundation of Ukraine, project 2020.01/0338 (M.K.)
and by the National Academy of Sciences of Ukraine, project KPKBK6541230 (Yu.H).}

\institutionalreview{Note applicable.}

\informedconsent{Note applicable.}

\dataavailability{Data for the plots of theoretical curves shown are available on request.}


\conflictsofinterest{The authors declare no conflict of interest.}

\appendixtitles{no} 
\appendixstart
\appendix
\section{}\label{app1}

In the Appendix, we evaluate integrals that enter formulas
(\ref{III.6}), (\ref{IV.2}). In particular, we are interested in
the behavior at small $\varepsilon$ of the following integrals:

\begin{equation}\label{A1}
I_\mu(\varepsilon)=\int_{\varepsilon}^{\infty}  dx\,
\frac{1}{x^\mu}\ln \cosh x ,
\end{equation}

\begin{equation}\label{A2}
I_{\lambda,\mu}(\varepsilon)=\int_{\varepsilon}^{\infty}  dx
\int_{\varepsilon}^{\infty}  dy\, \frac{1}{x^\lambda y^\mu}  \ln
\cosh (xy) ,
\end{equation}
We will consider the region where $\lambda, \mu >2$.

\subsection*{Integral $I_\mu(\varepsilon)$}

Let us first consider integral (\ref{A1}). At $2<\mu < 3$, it does
not diverge for $\varepsilon\to 0$, therefore its leading behavior
in this limit can be evaluated by numerical integration:
\begin{equation}\label{A2a}
I_\mu(\varepsilon)= i_\mu + O(\varepsilon),
\end{equation}
with
\begin{equation}\label{A2b}
i_\mu=\int_{0}^{\infty}  dx\, \frac{1}{x^\mu}\ln \cosh x  ,
\hspace{2em} 2<\mu <3\, .
\end{equation}
\textls[-15]{Numerical values of this and further constants $i_\mu$ are plotted as a function of $\mu$ in Figure~\ref{fig3}.} With a further increase in $\mu$, first,
the logarithmic singularity appears at $\mu=3$. It can be singled
out, leading to:
\begin{equation}\label{A2c}
I_3(\varepsilon)= - \frac{\ln \varepsilon}{2} + i_3 +
O(\varepsilon^2),
\end{equation}
where $i_3=0.64525$.

For $\mu>3$, to single out the leading singularities of the
function under integration at small $x$, we integrate twice by parts,
resulting in:
\begin{equation}\label{A2d}
I_\mu(\varepsilon)= -\frac{\varepsilon^{1-\mu}\ln \cosh
\varepsilon}{1-\mu} + \frac{\varepsilon^{2-\mu}\tanh
\varepsilon}{(1-\mu)(2-\mu)} -
\frac{\varepsilon^{3-\mu}}{(1-\mu)(2-\mu)(3-\mu)}  +
i_\mu(\varepsilon),
\end{equation}
with
\begin{equation}\label{A4}
i_\mu(\varepsilon)=\frac{1}{(\mu-1)(2-\mu)}\int_{\varepsilon}^{\infty}
dx\, x^{2-\mu}(\tanh x)^2 .
\end{equation}
Further analysis depends on the value of $\mu$. In the region
$3<\mu<5$, the integral on the right-hand side of Equation (\ref{A4})
converges at $\varepsilon \to 0$ and its leading asymptotics can be
evaluated numerically. So, keeping the leading behavior of the first
three terms in (\ref{A2d}) results in:
\begin{equation}\label{A6b}
I_\mu(\varepsilon)= \frac{\varepsilon^{3-\mu}}{2(\mu-3)} + i_\mu +
O(\varepsilon), \hspace{2em} 3<\mu<5,
\end{equation}
with
\begin{equation}\label{A6}
i_\mu=\frac{1}{(\mu-1)(2-\mu)}\int_{0}^{\infty}  dx\,
x^{2-\mu}(\tanh x)^2.
\end{equation}

Logarithmic singularity appears in (\ref{A4}) at $\mu=5$, leading to:
\begin{equation}\label{A7}
I_\mu(\varepsilon)=\varepsilon^{-2}/4 -(\ln \varepsilon)/12+i_5
+O(\varepsilon),
\end{equation}
with $i_5=-0.11309$.

\begin{figure}[H]
{\includegraphics[width=10cm]{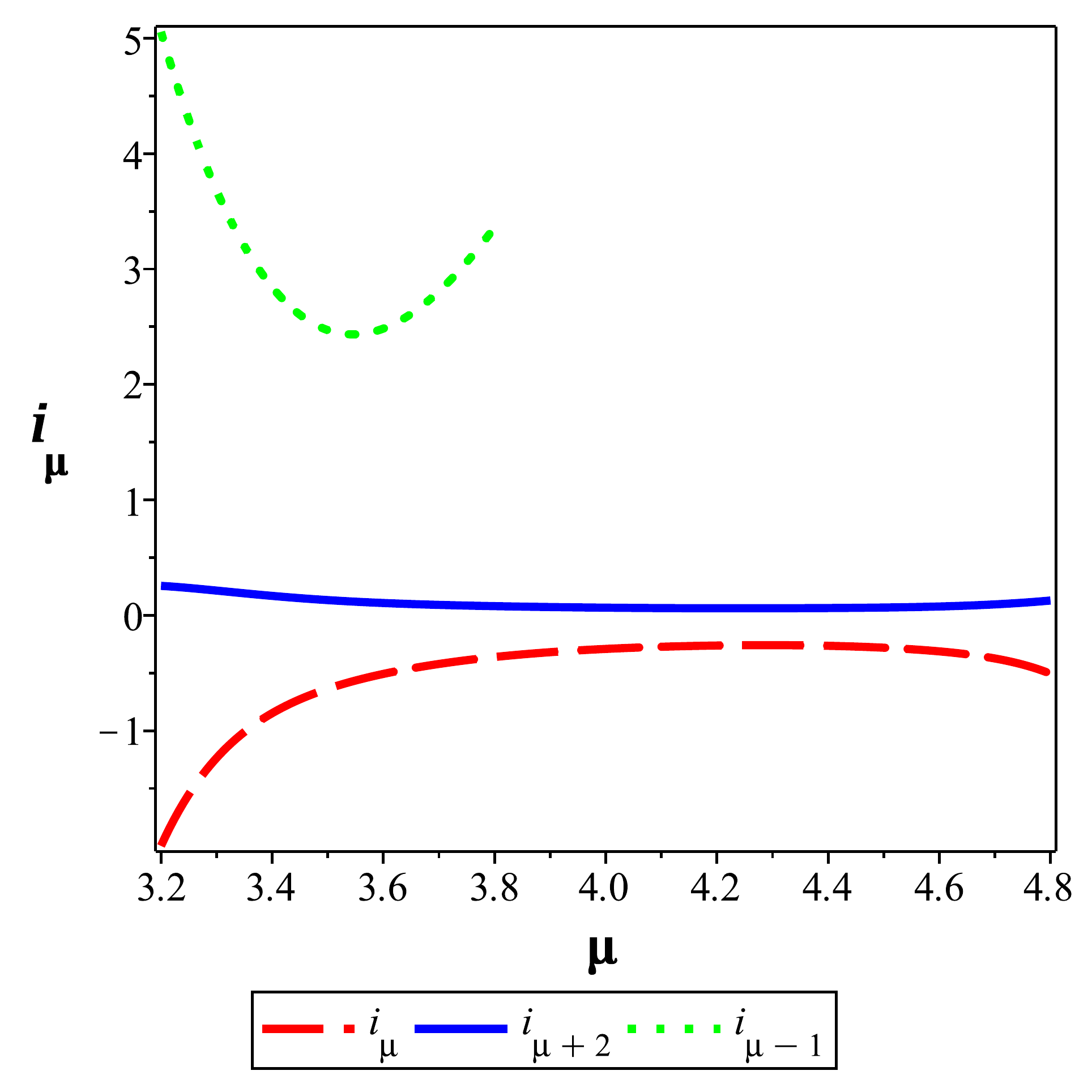}}
 \vspace*{8pt}
\caption{Dependence of constants $i_\mu$ in Equations (\ref{A2a}) and (\ref{A6b}),
(\ref{A9b}) on $\mu$.
  \label{fig3}}
\end{figure}

For higher values of $\mu$, analysis can be performed in a similar
fashion. In particular, for $5 < \mu < 7$, again integrating twice by
parts, one extracts a power-law singularity from the integral
(\ref{A4}):
\begin{equation}\label{A9b}
I_\mu(\varepsilon)= \frac{\varepsilon^{3-\mu}}{2(\mu-3)} -
\frac{\varepsilon^{5-\mu}}{12(\mu-5)} +  i_\mu +O(\varepsilon)\, ,
\hspace{2em} 5<\mu<7\, ,
\end{equation}
with
\begin{eqnarray}\nonumber
i_\mu &=&\frac{2}{(\mu-1)(2-\mu)(3-\mu)}\Big [ \int_{0}^{\infty}
dx\, x^{3-\mu}(\tanh x)^3 - \\ && \label{A9}
\frac{1}{4-\mu}\int_{0}^{\infty} dx\, x^{4-\mu}(\tanh x)^2\Big ] \,
.
\end{eqnarray}

Summarizing the above derived expressions for the leading behavior
of the integral (\ref{A1}) at small $\varepsilon$, we obtain the
following useful formula:

\begin{eqnarray}\label{A10b}
I_\mu(\varepsilon) - i_\mu \simeq\left\{
\begin{array}{lll}
& O(\varepsilon),  & 2 < \mu <3 \, , \\
& -(\ln \varepsilon)/2 + O(\varepsilon^2),  & \mu =3\, , \\
& \varepsilon^{3-\mu}/(2(\mu-3)) + O(\varepsilon),  & 3 < \mu < 5\, , \\
& \varepsilon^{-2}/4  -(\ln \varepsilon)/12 +O(\varepsilon),  & \mu = 5\, , \\
& \varepsilon^{3-\mu}/(2(\mu-3)) - \varepsilon^{5-\mu}/(12(\mu-5)) +
                O(\varepsilon),  & 5 < \mu < 7\, .
\end{array}
\right.
\end{eqnarray}

Constants $i_\mu$ for different $\mu$ can be evaluated numerically using formulas (\ref{A2b}),
(\ref{A6}), (\ref{A9}).
Their dependence on $\mu$ is shown in Figure~\ref{fig3}. They can be also checked against
analogous constants evaluated in Ref. \cite{Krasnytska16} using
different integral representations.

\subsection*{ Integral $I_{\lambda,\mu}(\varepsilon)$.  }

To single out leading singularities of the integral
$I_{\lambda,\mu}(\varepsilon)$  at small $\varepsilon$, we
differentiate Equation (\ref{A2}) with respect to $\varepsilon$. Due to
the fundamental theorem of calculus, the result~reads:
\begin{equation}\label{A11}
\frac{{\rm d}\, I_{\lambda,\mu}(\varepsilon)}{{\rm d}\, \varepsilon}
= -\varepsilon^{\lambda-\mu-1} I_\lambda(\varepsilon^2) -
\varepsilon^{\mu - \lambda-1} I_\mu(\varepsilon^2)\, ,
\end{equation}
where the asymptotic behavior of the integrals in the r.h.s. of Eq.
(\ref{A11}) is defined by Equation (\ref{A10b}) provided the substitution
$\varepsilon \to \varepsilon^2$. Consequently, the asymptotic
behavior of $J_{\lambda,\mu}(\varepsilon)$ is obtained by
integrating Equation (\ref{A11}) with respect to $\varepsilon$. In
particular, at $\lambda=\mu$, Equation (\ref{A11}) reduces to
\begin{equation}\label{A12}
\frac{{\rm d}\, I_{\lambda,\lambda}(\varepsilon)}{{\rm d}\,
\varepsilon} = -2\, \varepsilon^{-1} I_\lambda(\varepsilon^2)\, ,
\end{equation}
and one readily obtains:
\begin{eqnarray}\label{A13}
I_{\lambda,\lambda}(\varepsilon) - i_{\lambda,\lambda} + 2 i_\lambda
\ln \varepsilon\simeq  \left\{
\begin{array}{lll}
& O(\varepsilon^2),  & 2 < \lambda <3 \, , \\
&(\ln \varepsilon)^2/2 + O(\varepsilon^4),  & \lambda =3\, , \\
& \frac{\varepsilon^{6-2\lambda}}{2(\lambda-3)^2} + O(\varepsilon^2),  & 3 < \lambda < 5\, , \\
& \varepsilon^{-4}/8   + (\ln \varepsilon)^2/6 +O(\varepsilon^2),  & \lambda = 5\, , \\
& \frac{\varepsilon^{6-2\lambda}}{2(\lambda-3)^2} -
\frac{\varepsilon^{10-2\lambda}}{12(\lambda-5)^2}+ O(\varepsilon^2),
& 5 < \lambda < 7\, .
\end{array}
\right.
\end{eqnarray}
Constants $i_\lambda$ have been defined above and numerical values of the integration constants $i_{\lambda,\lambda}$ are not necessary for our analysis. 

Noting that integral (\ref{A2}) is symmetric with respect to
interchange of its indices,
$$
I_{\lambda,\mu}(\varepsilon) = I_{\mu,\lambda}(\varepsilon),
$$
it is enough to make a further evaluation in the region $\mu>\lambda$.
The resulting expressions read:
\begin{itemize}
    \item $2 < \lambda <3$:

\begin{eqnarray}\label{A13a}
I_{\lambda,\mu}(\varepsilon) - i_{\lambda,\mu} = \left\{
\begin{array}{lll}
& i_\lambda\frac{\varepsilon^{\lambda-\mu}}{\mu-\lambda}
+{O(\varepsilon^{\lambda-\mu+2})}, & 2 < \mu <3 \, , \\
& i_\lambda\frac{\varepsilon^{\lambda-3}}{3-\lambda}+\frac{\ln \varepsilon}{3-\lambda}\varepsilon^{{3-\lambda}}
+{O(\varepsilon^{\lambda-{1}})},  & \mu =3\, , \\
& i_\lambda\frac{\varepsilon^{\lambda-\mu}}{\mu-\lambda}-\frac{\varepsilon^{6-\lambda-\mu}}{2(6-\lambda-\mu)(\mu-3)}+O(\varepsilon^{\lambda-\mu+2}),  & 3 < \mu < 5\, , \\
& i_\lambda\frac{\varepsilon^{\lambda-5}}{5-\lambda}-\frac{\varepsilon^{1- \lambda}}{4(1-\lambda)}+O(\varepsilon^{\lambda-{3}}),  & \mu = 5\, , \\
&i_\lambda\frac{\varepsilon^{\lambda-\mu}}{\mu-\lambda}-
\frac{\varepsilon^{6-\lambda-\mu}}{2(6-\lambda-\mu)(\mu-3)}
+O(\varepsilon^{\lambda-\mu+2}), & 5 < \mu < 7\, .
\end{array}
\right.
\end{eqnarray}

    \item $\lambda =3$:
\begin{adjustwidth}{-4.6cm}{0cm}
    \begin{eqnarray}\label{A13b}
I_{\lambda,\mu}(\varepsilon) - i_{\lambda,\mu} = \left\{
\begin{array}{lll}
& \frac{\varepsilon^{3-\mu}}{3-\mu}\ln \varepsilon
-\varepsilon^{3-\mu}[\frac{i_3}{3-\mu}+\frac{1}{2(\mu-3)^2}]
+{O(\varepsilon^{\mu-1},\varepsilon^{7-\mu})},  & 3 < \mu < 5\, , \\
&  \varepsilon^{-2}[i_3/2 - 1/8]- \varepsilon^{-2} {\ln \varepsilon}/2
+{O(\varepsilon^{2})}, & \mu = 5\, , \\
& \frac{\varepsilon^{3-\mu}}{3-\mu}\ln \varepsilon
-\varepsilon^{3-\mu}[\frac{{i_3}}{3-\mu}+\frac{1}{2(\mu-3)^2}] 
+{O(\varepsilon^{7-\mu})}, & 5
< \mu < 7\, .
\end{array}
\right.
\end{eqnarray}
\end{adjustwidth}
    \item $3 < \lambda < 5$:

    \begin{eqnarray}\label{A13c}
    I_{\lambda,\mu}(\varepsilon) - i_{\lambda,\mu} = \left\{
\begin{array}{lll}
& i_\lambda\frac{\varepsilon^{\lambda-\mu}}{\mu-\lambda}{+}\frac{\varepsilon^{6-\lambda-\mu}}{2{(\lambda-3)(\mu-3)}}
+{O(\varepsilon^{\lambda-\mu+2})},  & 3 < \mu < 5\, , \\
& i_\lambda\frac{\varepsilon^{\lambda-5}}{5-\lambda}+\frac{\varepsilon^{1- \lambda}}{4(\lambda-3)} 
+{O(\varepsilon^{\lambda-3})},  & \mu = 5\, , \\
&i_\lambda\frac{\varepsilon^{\lambda-\mu}}{\mu-\lambda}+
\frac{\varepsilon^{6-\lambda-\mu}}{{2}(\lambda-3)(\mu-3)}- \\ &
\frac{\varepsilon^{10-\lambda-\mu}}{12(10-\lambda-\mu)(\mu-5)}
+O(\varepsilon^{\lambda-\mu+2}), & 5 < \mu < 7\, .
\end{array}
\right.
\end{eqnarray}

    \item $\lambda =5$, {$5 < \mu < 7$}:
\begin{adjustwidth}{-4.6cm}{0cm}
    \begin{equation} \label{A13d}
I_{\lambda,\mu}(\varepsilon) - i_{5,\mu} =
\frac{\varepsilon^{5-\mu}}{{6}(5-\mu)}\ln \varepsilon -
\varepsilon^{5-\mu}[\frac{i_5}{{5-\mu}}+\frac{1}{12(\mu-5)^2}]+\frac{\varepsilon^{1-\mu}}{4(\mu-3)} +{O(\varepsilon^{7-\mu})}.
\end{equation}
\end{adjustwidth}
    \item $5 < \lambda < 7$,  {$5 < \mu < 7$}:

\begin{equation}\label{A13e}
I_{\lambda,\mu}(\varepsilon) - i_{\lambda,\mu} =
i_\lambda\frac{\varepsilon^{\mu - \lambda}}{\lambda-\mu}+
\frac{\varepsilon^{6-\lambda-\mu}}{2{(\lambda-3)(\mu-3)}}-
\frac{\varepsilon^{10-\lambda-\mu}}{12{(\lambda-5)(\mu-5)}}
+{O(\varepsilon^{\lambda-\mu+2})}.
\end{equation}

\end{itemize}
Asymptotic estimates (\ref{A13a})--(\ref{A13e}) together with 
(\ref{A13}) are used in the study
to obtain expressions for the free energy of the model.

\end{paracol}
\reftitle{References}


\end{document}